\documentclass[a4paper,aps,onecolumn,showpacs,showkeys,nofootinbib,preprintnumbers,superscriptaddress,amsmath,amssymb,amsfonts]{revtex4}
\usepackage{graphicx}
\usepackage{bm,natbib}
\usepackage{amsmath}
\usepackage[english]{babel}
\usepackage[T1]{fontenc}
\usepackage{amssymb}
\usepackage{amsfonts}
\usepackage{epsfig}
\usepackage{colordvi}
\usepackage{psfrag}
\usepackage{color}
\usepackage{ mathrsfs }
\newcommand{\La}{\mathscr{L}}
\newcommand{\Lagr}{\mathcal{L}}
\newcommand{\G}{\mathcal{G}}

\usepackage{dcolumn}
\usepackage{multirow}
\usepackage{hyperref}
\usepackage{epstopdf}
\usepackage{bm}
\usepackage{times}
\usepackage{float}
\restylefloat{table}

\hypersetup{
  colorlinks=true,        
  linkcolor=blue,         
  citecolor=magenta,      
}

\begin{document}
\title{Higher dimensional static and spherically symmetric solutions in extended Gauss-Bonnet gravity}

\author{Francesco Bajardi}
\email{bajardi@na.infn.it}
\affiliation{Department of Physics ``E. Pancini'', University of Naples ``Federico II'', Naples, Italy,}
\affiliation{INFN Sez. di Napoli, Compl. Univ. di Monte S. Angelo, Edificio G, Via	Cinthia, I-80126, Napoli, Italy.}

\author{Konstantinos F. Dialektopoulos}
\email{kdialekt@gmail.com}
\affiliation{Center for Gravitation and Cosmology, College of Physical Science and Technology, Yangzhou University, Yangzhou 225009, China}

\author{Salvatore Capozziello}
\email{capozziello@na.infn.it}
\affiliation{Department of Physics ``E. Pancini'', University of Naples ``Federico II'', Naples, Italy,}
\affiliation{INFN Sez. di Napoli, Compl. Univ. di Monte S. Angelo, Edificio G, Via	Cinthia, I-80126, Napoli, Italy.}
\affiliation{Gran Sasso Science Institute, viale F. Crispi 7, I-67100, L'Aquila, Italy.}
\affiliation{Tomsk State Pedagogical University, ul. Kievskaya, 60, 634061 Tomsk, Russia.}

\begin{abstract}
We study a theory of gravity  of the form $f(\mathcal{G})$ where $\mathcal{G}$ is the Gauss-Bonnet topological invariant without considering the standard Einstein-Hilbert term as common in the literature, in arbitrary $(d+1)$ dimensions.  The approach is motivated by the fact that, in particular conditions, the Ricci curvature scalar can be easily recovered and then a pure $f(\cal G)$ gravity can be considered a further generalization of General Relativity like $f(R)$ gravity. Searching for Noether symmetries,  we specify the functional forms  invariant under point transformations in a static and spherically symmetric spacetime and, with the help of these symmetries, we find exact solutions showing that Gauss-Bonnet gravity is significant without assuming the Ricci scalar in the action.  
\end{abstract}



\maketitle

\section{Introduction}
\label{sec:intro}

Even though Einstein's General Relativity (GR) and the related cosmological model, $\Lambda$CDM, have been  successful according to a wide range of observations (Solar System tests, supernova type Ia, large scale structure, cosmic microwave background and more), there are some shortcomings that have to be addressed in view of a final theory of gravity and a self-consistent cosmological model \cite{Perivolaropoulos:2008ud,Bull:2015stt,Barack:2018yly}. The nature of the ``dark sector'' of the Universe, \textit{i.e.} dark matter and dark energy, the huge discrepancy between the theoretical value of the cosmological constant with the observed one, as well as the inability to find a TeV-scale supersymmetry are some of the puzzles of modern physics. These issues, together with the unknown quantum nature of gravitational interaction,  singularities,  coincidence problem in cosmology and more, initiated the hunt for an alternative description of gravity (see e.g. \cite{Capozziello:2011et,Nojiri:2010wj,Clifton:2011jh,Nojiri:2017ncd}, and references therein).

During the last two decades, there have been several  approaches aimed  to find out a more general description for the gravitational interaction. Generalization of the Einstein-Hilbert action, like $f(R)$ theories \cite{Capozziello:2002rd,Capozziello:2009nq}, addition of extra fields, like Brans-Dicke theory \cite{Brans:1961sx}, Horndeski theory \cite{Horndeski,Deffayet:2009wt,Capozziello:2018gms}, Tensor-Vector-Scalar theory, bimetric theories \cite{Hassan:2011zd}, massive gravity \cite{deRham:2010kj}, non-local theories \cite{Deser:2007jk,Modesto:2017sdr}, higher dimensional gravitational theories in the framework of tangent Lorentz bundles \cite{Ikeda:2019ckp}, as well as reformulations such as the Teleparallel Equivalent of General Relativity (TEGR) \cite{Perreirabook} and its modifications \cite{Cai:2015emx}, are some of the approaches studied in detail, not only at cosmological scales, but also at astrophysical ones.

On the other hand, in the quantum regime, there have been many attempts to find a quantum formulation of gravity leading to higher dimensional theories, like Kaluza-Klein model, DGP model, Einstein-Dilaton-Gauss-Bonnet gravity, as well as generalizations such as the Lovelock gravity \cite{Lovelock:1971yv}. In the low energy effective action of string/M-theory, a specific curvature  invariant  naturally appears. It is  Gauss-Bonnet (GB) scalar \cite{Gasperini:1992em} given by
\begin{equation}
\label{GBterm}
\G = R^2 - 4 R_{\mu\nu}R^{\mu\nu}+R_{\alpha\beta\mu\nu}R^{\alpha\beta\mu\nu}\,,
\end{equation} 
where $R,\,R_{\mu\nu}$ and $R_{\alpha\beta\mu\nu}$ are the Ricci scalar, the Ricci tensor and the Riemann tensor respectively. This term is a topological invariant in $3+1$ dimensions (or less). Practically, this means that a linear term in $\G$, in the Einstein-Hilbert action, would not affect the equations of motion. However, in the literature, an addition of an arbitrary function $f(\G)$ has been proposed \cite{Nojiri:2005jg}. Specifically, the theory given by the action $\mathcal{S} = \int d^4x\sqrt{-g} \left[\frac{1}{2\kappa^2}R + f(\G) \right]$ has been extensively studied. In \cite{DeFelice:2008wz}, cosmologically viable models are considered  by studying the stability of a late-time de-Sitter solution and the existence of radiation and matter epochs. In \cite{Uddin:2009wp},  possible power-law scaling solutions have been taken into account by developing the scalar tensor equivalent of the above theory. In particular, in \cite{DeFelice:2009rw},  authors  study cosmological perturbations and show that density perturbations cause instabilities. In \cite{Davis:2007ta}, the author shows that the above theory is ruled out as a possible explanation of the late-time acceleration by Solar System tests. In \cite{Chakraborty:2015taq}, the Gauss-Bonnet term is added to a $f(R)$ five-dimensional Lagrangian and a static spherically symmetric solution is studied. In \cite{deBoer:2009pn} the authors study the energy bounds for Gauss-Bonnet gravity in an $AdS_7$ background. Finally in \cite{Astashenok:2015haa},  a mimetic version of the above theory is considered and they find, besides  accelerating behaviors, solutions that unify the inflation era together with dark energy. In addition,  dark matter can be  described in the framework of this model.  In \cite{felix}, the Newtonian and Post-Newtonian limit of these models is studied in detail.

For almost half a century, higher dimensional theories of gravity have been studied in many different contexts in the literature \cite{Rubakov:1983bb}. The aforementioned puzzling phenomena in gravity can sometimes be explained by invoking extra dimensions \cite{Antoniadis:1990ew,Randall:1999ee,Randall:1999vf,ArkaniHamed:1998rs}. Braneworlds and other higher dimensional modifications of Einstein's theory, e.g. Lovelock theory \cite{Lovelock:1971yv} have been considered as possible extensions in the hunt for a self-consistent  theory of gravity. 

All of the above researches  deal with a theory that safely recovers GR in the background or in some limit. This means that if one switches off the effect of the GB contribution, i.e. $f(\G)\rightarrow 0$, then the action reduces to the Einstein-Hilbert and one recovers GR. This happens because  GR has to be restored in view of  observations  and experimental tests.  In this paper, we propose a  scenario where GR is not in the background "a priori" and gravity is given only by quadratic curvature invariants and specifically by an arbitrary function of the GB term. However, GR can be restored as a particular case of $f(\cal G)$ gravity  and the further degrees of freedom related to $R_{\mu\nu}$ and
 $R^{\alpha}{}_{\lambda\mu\nu}$ can be neglected with respect to $R$. This happens if particular symmetries are adopted like in homogeneous and isotropic cosmology or in other specific cases.

Here, we consider  a spherically symmetric background and search for  Noether Symmetries in general $(d+1)$ dimensions. Specifically, we use the so called {\it Noether Symmetry Approach} \cite{Dialektopoulos:2018qoe,Capozziello:1996bi}, which has been extensively used in the literature as a geometric criterion to select forms of the arbitrary functions  in several alternative gravity theories that are invariant under point transformations (see for example  
 \cite{Dialektopoulos:2018qoe,Basilakos:2013rua,Bahamonde:2016grb,Capozziello:2016eaz,Bahamonde:2017sdo}).

Here, we adopt the above approach for $f(\G)$ gravity in spherical symmetry for arbitrary $(d+1)$ dimensions.  It is interesting to point out that, the only  forms of $f(\G)$ selected by symmetries   are  power-law functions.  By  these symmetries, it is possible to  find exact spherically symmetric solutions, 
 which for specific values of the power-law,  provide the same prediction as GR. I means that standard GR can be recovered from $f(\G)$ gravity.


This paper is organized as follows: in Sec. \ref{secII}, we present $f(\G)$ gravity  and derive the field  equations. Furthermore, we construct the point-like Lagrangian in a spherically symmetric minisuperspace. Sec. \ref{secIII} is devoted to  the Noether Symmetry Approach and,  in Sec. \ref{secIV},  we use it to find the forms of $f(\G)$ selected by the Noether Symmetries. Moreover, in Sec. \ref{secIV}, we find  exact spherically symmetric solutions by making use of the symmetries. 
Finally, in Sec.\ref{secV},  we draw conclusions  and discuss future perspectives. Throughout the paper the metric signature is $\left(+---\right)$ and physical units $\hbar = c = k_B = 8\pi G = 1$ are adopted.

\section{The Gauss-Bonnet gravity in spherical symmetry}
\label{secII}

A general Gauss-Bonnet gravity theory is given by the action
\begin{equation}\label{f(G)}
S = \int \sqrt{|g|} f(\G) \; d^{d+1} x\,,
\end{equation}
where $\G$ is the Gauss-Bonnet invariant  given by Eq. \eqref{GBterm}. 
In four dimensions (\textit{i.e.} $d=3$), a  linear term  $\G$ in the action is trivial because, as a topological invariant, it turns into a surface term and the related integral is null. As already mentioned in the Sec. \ref{sec:intro}, up to now, people studied $f(\G)$ theories in $d=3$, adding  a Ricci scalar in the action \eqref{f(G)}, in order to recover General Relativity for  $f(\G)\rightarrow 0$. In our case, we consider pure $f(\G)$ theories and we claim that GR can be recovered without considering the Einstein-Hilbert term \textit{a priori} in the action. 

By varying \eqref{f(G)} with respect to the metric, we get the field equations
\begin{align}\label{feq}
\frac{1}{2} g_{\mu \nu} f &- \left(2R R_{\mu \nu} - 4 R_{\mu \alpha} R^{\alpha}{}_\nu + 2 R_\mu {}^{\alpha \beta \gamma} R_{\nu \alpha \beta \gamma} - 4 R^{\alpha \beta} R_{\mu \alpha \nu \beta}\right) f_\G +\nonumber
\\
&+ \left[2R \nabla_\mu \nabla_\nu +4 G_{\mu \nu} \Box - 4 (R^{\rho}{}_{ \nu} \nabla_{\mu} +R^{\rho}{}_{ \mu} \nabla_{\nu}) \nabla_{\rho} + 4 g_{\mu \nu} R^{\rho \sigma} \nabla_\rho \nabla_\sigma - 4 R_{\mu \alpha \nu \beta} \nabla^\alpha \nabla^\beta \right] f_\G = 0
\end{align}
where $f_{\G}$ is the derivative of $f$ with respect to $\G$. It is worth using also   the trace of Eq. \eqref{feq}, that is 
\begin{equation}\label{treq}
\frac{d+1}{2}f - 2 \G f_{\G}-2(d-2) \left(R\square  - 2 R^{\mu\nu}\nabla_{\mu}\nabla_{\nu}\right)f_{\G} = 0\,.
\end{equation}
This can be seen as the equation of motion for the new scalar degree of freedom introduced in this theory. It is already known \cite{Kobayashi:2011nu} that, the theory \eqref{f(G)} in $d=3$ is a part of the Horndeski action and thus contains an extra scalar degree of freedom. 

\subsection{Spherical Symmetry}
Let us  consider now a static and spherically symmetric \textit{Ansatz} for the metric, that reads
\begin{equation}\label{sphsymmet}
ds^2 = P(r)^2 dt^2 - Q(r)^2 dr^2 - r^2 d\Omega_{d-1}^2\,, 
\end{equation}
where $d\Omega_{d-1}^2 = \sum_{j=1}^{d-1} d\theta_j^2 + \sin^2 \theta_j d\phi^2$ is the metric element of the $(d-1)$-sphere for a spacetime labeled by coordinates $x^\mu = (t,r,\theta_1,\theta_2,..., \theta_{d-2}, \phi)$. Before  proceeding, an important comment is necessary here; we assume that the metric \eqref{sphsymmet} is not dynamical, which means that the Birkhoff's theorem should be valid for these models. This is not proven and we take it for granted in theories such as \eqref{f(G)}. However, there are a lot of references in the literature claiming to have found cases where a generalization of the Birkhoff's theorem could exist \cite{Bogdanos:2009pc,Deppe:2012wk,Bogdanos:2010zz,Dotti:2010bw,Zegers:2005vx,Ray:2015ava}.

The Gauss-Bonnet term \eqref{GBterm} in arbitrary $(d+1)$ dimensions takes the form
\begin{align}
\label{GBterm_spherical}
\G = \frac{(d-1) (d-2)}{r^4} \Big[ 4 r^2 P (P^2 - 1) P'' + 8 (d-3) r P (P^2 - 1) P' + 4 r^2 &(3 P^2-1) P'^2 +\nonumber\\
&+ (d-3)(d-4) (P^4 - 2P^2 +1) \Big]\,,
\end{align}
where the prime stands for derivatives with respect to the radial coordinate and we set for simplicity $\theta_j = \pi/2$. Note that for $ d \le 2$ (\emph{i.e} in less than four dimensions), the above scalar vanishes identically, while for $d=3$, it becomes a topological surface term.

In order to calculate the point-like Lagrangian of the  theory for \eqref{sphsymmet}, we introduce a  Lagrange multiplier as \cite{Capozziello:2007wc,Bahamonde:2017sdo,Capozziello:2016eaz}
\begin{equation}
\mathcal{S} = \int d^{d+1} x \; r^{d-1} PQ \left[f(\G) - \lambda \left(\G - \tilde{\G} \right) \right]\,,
\end{equation}
with $\tilde{\G}$ being the Gauss-Bonnet term in spherical symmetry \eqref{GBterm_spherical} and $\lambda$ the Lagrange multiplier given by varying the action with respect to $\G$, i.e. $\lambda =\partial _{\G} f$. Substituting $\tilde{\G}$ and integrating out the second derivatives,  we obtain 
\begin{align}
\label{pointlikeLagra}
\La(r,P,Q,\G)=  &r^{d-1} P Q \left[f - \G f_{\G}\right] + \nonumber
\\ \nonumber
\\
&+ \frac{(d-1)(d-2) r^{d-5} (Q^2-1)}{Q^4} \left\{(d-3) P f_\G \left[ (d-4)Q (Q^2-1)+ 4 r Q' \right] + 4 r^2 Q P' \G' f_{\G\G} \right\}\,,
\end{align} 
where $f_\G$ and $f_{\G\G}$ are the first and second derivatives of $f$ with respect to $\G$. This is the point-like and canonical Lagrangian of our theory in a static and spherically symmetric spacetime. Its configuration space is $\mathcal{Q} = \{P,Q,\G\}$, and the tangent space $\mathcal{TQ} = \{P,P',Q,Q',\G,\G'\}$. 

\section{The Noether Symmetry Approach}
\label{secIII}

Let us briefly introduce some basic notions  of the so called Noether Symmetry Approach \cite{Dialektopoulos:2018qoe,Capozziello:1996bi}. 
Noether symmetries are a subclass of Lie point symmetries applied in dynamical systems described by a Lagrangian density. Noether's theorem states that if 
\begin{equation}
X = \xi \partial _t + \eta^i\partial _{q^i}\,,
\end{equation}
is a generator of infinitesimal point transformations, then the Lagrangian density is invariant under $X$ if and only if
\begin{equation}
X^{[1]}\Lagr + \dot{\xi}\Lagr = \dot{g}\,,\label{Teorema}
\end{equation}
with $g$ being a function of the affine parameter $t$ and the generalized coordinates $q^i$ and $X^{[1]}$ is the first prolongation of $X$.

The $n$-prolongation of the generator has the form
\begin{equation}
X^{[n]} = \xi \frac{\partial }{\partial t} + \eta^i \frac{\partial }{\partial q^i} + \eta^{i \; [1]} \frac{\partial}{\partial \dot{q}^i} + ... + \eta^{i \; [n]} \frac{\partial}{\partial \frac{d^n q^i}{dt^n}}\,,
\end{equation}
with
\begin{equation}
\eta^{i \; [n]} = \frac{d \eta^{i \; [n-1]}}{dt} - \dot{\xi} \frac{d^n q^i}{dt^n}\,.
\end{equation}
The parameter $t$ represents any affine parameter and it is chosen depending on the symmetry of the spacetime. Then we have
\begin{equation}
X^{[1]} = \xi \frac{\partial }{\partial t} + \eta^i \frac{\partial }{\partial q^i} + \eta^{i \; [1]} \frac{\partial}{\partial \dot{q}^i} = \xi \frac{\partial }{\partial t} + \eta^i \frac{\partial }{\partial q^i} + (\dot{\eta}^i - \dot{q}^i \dot{\xi}) \frac{\partial}{\partial \dot{q}^i} \,.\label{prolungamento}
\end{equation}

It is easy to extend the above to a general Lagrangian density that depends on $x^{\mu}$ parameters. Specifically, the prolongation \eqref{prolungamento} becomes
\begin{equation}
X^{[1]} = \xi^\mu \partial_\mu + \eta^i \frac{\partial }{\partial q^i} + (\partial_\mu \eta^i - \partial_\mu q^i \partial_\nu \xi^\nu) \frac{\partial}{\partial (\partial_\mu q^i)} \label{prolungamento generalizzato}
\end{equation}
and the Noether's theorem \eqref{Teorema}
\begin{equation}
X^{[1]} \La + \partial_\mu \xi^\mu \La = \partial_\mu g^\mu \label{teorema generalizzato}\,.
\end{equation}

In more details, let us consider the following transformation
\begin{equation}
\La( x^\mu, q^i, \partial_\mu q^i) \to \La( \tilde{x}^\mu, \tilde{q}^i, \partial_\mu \tilde{q}^i) \,,
\end{equation}
where transformation of $x^\mu$ and $q^i$ are given by:
\begin{equation}
\begin{cases}
\tilde{x}^\mu = x^\mu + \epsilon \xi^\mu(x^\mu, q^i) + O(\epsilon^2)\,,
\\
\tilde{q}^i = q^i + \epsilon \eta^i(x^\mu, q^i) + O(\epsilon^2)\,.
\end{cases}
\label{trasformazioni coordinate}
\end{equation}
The derivatives of the generalized coordinates $q^i$ transform as
\begin{equation}
\frac{d\tilde{q}^i}{d\tilde{x}^\mu} = \displaystyle \frac{dq^i + \epsilon d\eta^i}{dx^\mu + \epsilon d \xi^\mu} = \left( \frac{dq^i}{dx^\mu} + \epsilon \frac{d \eta^i}{dx^\mu} \right)\left(1 + \epsilon \frac{d\xi^\nu}{dx^\nu} \right)^{-1} \sim \left( \frac{dq^i}{dx^\mu} + \epsilon \frac{d \eta^i}{dx^\mu} \right)\left(1 - \epsilon \frac{d\xi^\nu}{dx^\nu} \right)\,,
\end{equation}
which at first order in $\epsilon$ has the form
\begin{equation}
\frac{d\tilde{q}^i}{d\tilde{x}^\mu} = \frac{dq^i}{dx^\mu} + \epsilon \left(\frac{d \eta^i}{dx^\mu} - \frac{dq^i}{dx^\mu} \frac{d \xi^\nu}{dx^\nu} \right) + O(\epsilon^2) = \partial_\mu q^i + \epsilon \left(\partial_\mu \eta^i - \partial_\mu q^i \partial_\nu \xi^\nu \right) + O(\epsilon^2)  \label{trasformazione derivata}\,.
\end{equation}
From Eq. \eqref{trasformazioni coordinate} and Eq. \eqref{trasformazione derivata} we can construct the generator of these transformations, that reads
\begin{equation}
X = \xi^\mu \partial_\mu + \eta^i \frac{\partial}{\partial q^i} \,.
\end{equation}
Now, if the transformations \eqref{trasformazioni coordinate} and \eqref{trasformazione derivata} hold,  the equations of motion, \textit{i.e.} the Euler-Lagrange equations, are invariant, and thus there exists a function $ g^\mu = g^\mu (x^\mu, q^i)$ such that the following condition holds
\begin{equation}
\frac{d \tilde{x}^\mu}{dx^\mu} \tilde{\La} = \La + \epsilon \partial_\mu g^\mu\,.
\end{equation}
The derivative with respect to $\epsilon$ will give
\begin{equation}
\La \frac{\partial}{\partial \epsilon} \frac{d \tilde{x}^\mu}{dx^\mu} + \frac{d \tilde{x}^\mu}{dx^\mu} \frac{\partial \tilde{\La}}{\partial \epsilon}  = \partial_\mu g^\mu\,,
\label{rel}
\end{equation}
and the transformations \eqref{trasformazioni coordinate} allow us to calculate the various terms. That is,
\begin{gather}
\frac{d \tilde{x}^\mu}{dx^\mu} = \frac{\partial \tilde{x^\mu}}{\partial x^\mu} + \frac{\partial \tilde{x^\mu}}{\partial q^i} \partial_\mu q^i = 1 + \epsilon \frac{\partial \xi^\mu}{\partial q^i} \partial_\mu q^i\,,
\label{rel1} \\
\frac{\partial}{\partial \epsilon} \frac{d \tilde{x}^\mu}{dx^\mu} = \frac{d}{dx^\mu}\left(\frac{\partial \tilde{x}^\mu}{\partial \epsilon}\right) = \partial_\mu \xi^\mu\,,
\label{rel2} \\
\frac{\partial \tilde{\La}}{\partial \epsilon} = \frac{\partial \tilde{\La}}{\partial \tilde{x}^\mu} \frac{\partial \tilde{x}^\mu}{\partial \epsilon} + \frac{\partial \tilde{\La}}{\partial \tilde{q}^i} \frac{\partial \tilde{q}^i}{\partial \epsilon} + \frac{\partial \tilde{\La}}{\partial (\partial_\mu \tilde{q}^i)} \frac{\partial (\partial_\mu \tilde{q}^i)}{\partial \epsilon}\,. 
\label{rel3}
\end{gather}
With the help of \eqref{trasformazione derivata},  we can replace \eqref{rel1}, \eqref{rel2} and \eqref{rel3} into \eqref{rel} and obtain
\begin{equation}
\left[\xi^\mu \partial_\mu + \eta^i \frac{\partial }{\partial q^i} + (\partial_\mu \eta^i - \partial_\mu q^i \partial_\nu \xi^\nu) \frac{\partial}{\partial (\partial_\mu q^i)} + \partial_\mu \xi^\mu \right] \La = \partial_\mu g^\mu\,,
\end{equation}
that is nothing else but \eqref{teorema generalizzato}. 
It is worth noticing that the associated Noether integral, which is the conserved quantity, is given by
\begin{equation}
\label{int2}
j^\mu = -\frac{\partial \La}{\partial (\partial_\mu q^i)} \eta^i + \frac{\partial \La}{\partial (\partial_\mu q^i)} \partial_\nu q^i \, \xi^\nu - \La \xi^\mu + g^\mu\,.
\end{equation}

In particular, for spherical symmetry, where the metric only depends on $r$, Eqs. \eqref{teorema generalizzato} and \eqref{prolungamento generalizzato} acquire the form:
\begin{equation}
X^{[1]} = \xi(r,q^i) \partial_r + \eta^i(r,q^i) \frac{\partial }{\partial q^i} + [\partial_r \eta^i(r,q^i) - \partial_r q^i \partial_r \xi(r,q^i)] \frac{\partial}{\partial (\partial_r q^i)} \label{prolungamento con r}\,,
\end{equation}
\begin{equation}
X^{[1]} \La + \partial_r \xi(r,q^i) \La = \partial_r g(r,q^i) \label{teorema r}\,.
\end{equation}
With this considerations in mind, let us apply the Noether Symmetry Approach to the point-like Lagrangian \eqref{pointlikeLagra}.


\section{Noether symmetries in Gauss-Bonnet gravity}
\label{secIV}

The generator of the point transformations \eqref{trasformazioni coordinate}, in our case, is given by
\begin{equation}
X = \xi(r,\G,P,Q) \partial_r +\eta^{\G} (r,\G,P,Q) \partial _{\G} + \eta^P(r,\G,P,Q) \partial_P + \eta^Q(r,\G,P,Q) \partial_Q\,,
\end{equation}
where $\xi$ and $\eta^i$, with $i = \{\G,P,Q\}$, are the components of  vector $X$.  By applying the Noether theorem, Eq. \eqref{teorema generalizzato}, we obtain a system of twelve equations which are not all independent. There are two non-trivial  functions $f(\G)$ determined by symmetries.
\begin{itemize}
\item
\textbf{Case 1:} In $d \geq 3$ dimensions,  we have $f(\G) = f_0 \G^n$ with $n\neq 1$.
The Noether symmetry of this model is given by
\begin{equation}\label{sym1}
X = c_1 r \partial _r - 4 c_1 \G \partial _{\G} + (4n - d) c_1 P \partial _P \,.
\end{equation}
and $g = c_2$, with $c_1$ and $c_2$ being constants. The invariant quantity \eqref{int2},  related to the above symmetry \eqref{sym1},  is 
\begin{align}
I = &\frac{1}{Q^3}c_1 f_0 r^{d-4} \G^{n-2} \Bigg[(1-n) r^4 \G^2 P Q^4 - 4 n(n-1)(d-2)(d-1) r \left(Q^2-1\right) \left(r P' + (d-4n) P \right) \G' -\nonumber \\
& - (d-2) (d-1)n \G  \left(Q^2-1\right) \Big[16 (n-1) r P' - (d-4) (d-3) P \left(Q^2-1\right)\Big]  \Bigg]-c_2\,.
\end{align}
\item
\textbf{Case 2:} In $d=4$ dimensions,  there is also the possibility of having a linear model of the form $f(\G) = f_0 \G$. Its Noether symmetry reads
\begin{equation}\label{sym2}
X = c_1 r \partial _r + c_2 \partial _P \,,
\end{equation}
and $g= c_3 - \frac{8 f_0 c_2 (3 Q^2-1)}{Q^3}$, with $c_1, c_2$ and $c_3$ being constants. Respectively, the preserved quantity related to \eqref{sym2} is
\begin{equation}
I = - \frac{8 f_0}{Q^3}\left(\frac{6 r \left(Q^2-3\right) \left(c_1 r P'-c_2\right) Q'}{Q}+c_2 \left(3 Q^2-5\right)\right) - c_3\,.
\end{equation}
\end{itemize}

\subsection{Spherically Symmetric solutions}
From the Noether theorem \eqref{teorema generalizzato},  we can build the following Lagrange system
\begin{equation}
\frac{dt}{\xi} = \frac{dq^i}{\eta^i} = \frac{d\dot{q}^i}{\eta^{[1],i}}\,.
\end{equation}
From the above system applied in our case, \textit{i.e.} for the symmetry \eqref{sym1},  we get the zero and first order invariants that read respectively
\begin{gather}
W^{[0],\G}(r,\G) = \frac{dr}{c_1 r}-\frac{d\G}{-4c_1\G}= \G r^4\,,\\W^{[0],P}(r,P) = \frac{dr}{c_1 r}-\frac{dP}{(4n-d)c_1P}=Pr^{d-4n}\,,\\
W^{[1],\G}(r,\G) = \frac{dr}{c_1 r}-\frac{d\G}{-4c_1\G}-\frac{d\G '}{-5 c_1 \G '}=\G ' r^{5}\,,\\
W^{[1],P}(r,P) = \frac{dr}{c_1 r}-\frac{dP}{(4n-d)c_1P}-\frac{dP'}{(4n-d-1)c_1P'}=P'r^{1+d-4n}\,.
\end{gather}
Using these, we can reduce the order of the equations of motion, from second to first and solve them.

The Lagrangian \eqref{pointlikeLagra} for the \textbf{Case 1} of the previous section, i.e. $f=f_0 \G^n$, becomes 
\begin{align}
\La = \frac{1}{Q^4}f_0 r^{d-5} \G^{n-2}& \Bigg\{\G P Q \Big[(d-4) (d-3) (d-2) (d-1) n \left(Q^4-2 Q^2+1\right)-\G (n-1) Q^4 r^4\Big] +\nonumber \\
&+4 (d-1) (d-2) n \left(Q^2-1\right) r \Big[(d-3) \G P Q'+(n-1) r Q \G' P'\Big]\Bigg\} 
\end{align}
and the associated Euler-Lagrange equations $\displaystyle \frac{\partial \La}{\partial q^i} = \frac{d}{dr} \frac{\partial \La}{\partial q'^i}$ are 
\begin{align}\label{ELP}
\frac{1}{Q^4}&f_0 r^{d-5} \G ^n\Bigg\{\frac{(d-2) (d-1) n \left(Q^2-1\right) \Big[(d-3) \G ^2 \left((d-4) Q \left(Q^2-1\right)+4 r Q'\right)-4 (n-1)(n-2) r^2 Q \G '^2\Big]}{\G ^3}-\nonumber \\
&-(n-1) r^4 Q^5+4 (d-2) (d-1) (n-1) n r\frac{ \G ' \Big[\left(Q^2-3\right) r Q'-(d-3) Q \left(Q^2-1\right)\Big]-Q \left(Q^2-1\right) r \G ''}{\G ^2}\Bigg\} = 0\,,\\ \label{ELQ}
\frac{1}{Q^4}&f_0 r^{d-5} \G ^{n-2}\Bigg\{ \G ^2 (n-1) P Q^4 r^4 +(d-2) (d-1) n \Big[4 (n-1) r \G ' \left((d-3) P \left(Q^2-1\right)+\left(Q^2-3\right) r P'\right)+ \nonumber \\
&+(d-3) \G \left(Q^2-1\right) \left(4 r P'-(d-4) P \left(Q^2-1\right)\right)\Big] \Bigg\} = 0 \,, \\ \label{ELG}
\frac{1}{Q^4}&f_0 (n-1) n r^{d-5} \G ^{n-2}\Bigg\{-\G P Q^5 r^4+(d-2) (d-1) \Big[(d-4) (d-3) P Q \left(Q^2-1\right)^2-\nonumber \\
&-4 r \left(\left(Q^2-1\right) \left(Q r P''-(d-3) P Q'\right)+P' \left((d-3) Q \left(Q^2-1\right)-\left(Q^2-3\right) r Q'\right)\right)\Big] \Bigg\} = 0\,,
\end{align}
for $P,\,Q$ and $\G$ respectively.
Solving Eq. \eqref{ELG} for $\G(r)$ we find, as expected, that $\G(r) = \tilde{\G}$, given by \eqref{GBterm_spherical}. We simplify Eq. \eqref{ELP} and Eq. \eqref{ELQ} by setting $Q(r) = 1/P(r)$ and we end up with one equation of the form
\begin{align}
(d-1)(d-2)\tilde{\G}^n\Bigg\{(d-3)(P^2 - 1)&\left[ (d-4)(P^2-1) - 4(n-2)r PP'\right] -\nonumber \\
&- 4(n-1)r^2\left[(P^2-1)PP''+(3P^2-1)P'^2 \right]\Bigg\} = 0\,. \label{EL1}
\end{align}
Obviously, for $d=1,2$ Eq. \eqref{EL1} is satisfied automatically. The rest of the equation accepts three solutions which read
\begin{gather}
P(r)^2 = 1 + e^{-2 k_2} \sqrt{k_1-4 r} \, r^{\frac{3}{2}-\frac{d}{2}} \label{firstND} \,\, \text{and} \,\, \G(r) = 0\, \quad n \neq 1 \quad d \ge 3, \\ \nonumber \\ 
P(r)^2 = P_0^2 \left(1- \frac{k_3}{r^{\frac{d}{2}-2}} \right) \,\, \text{and} \,\, \G(r) = 0\, \quad n = 1 \quad d \ge 4 \,, \label{secondND} \\ \nonumber \\
P(r)^2 = 1 \pm r^{2-\frac{d}{2}}\sqrt{\frac{4 k_1 d}{120 \binom{d+1}{d-4} }} \pm r^2 \sqrt{ \frac{ \G_0 (d-3) }{120 \binom{d+1}{d-4} }} \, \,\, \text{and} \,\, \G(r) = \G_0, \quad f(\G) = f_0 \G^{\frac{d+1}{4}}\quad d \ge 4\,, \label{lastND}
\end{gather}
with $k_1, k_2, k_3, P_0, \G_0$ constants. These are general black hole solutions for the theory \eqref{f(G)} with $f(\G) = f_0 \G ^n$; in particular, the first one is valid in arbitrary $d$ dimension with $d \ge 3$, while the other holds in more than four dimensions. Solution \eqref{lastND}, which is the (A)dS equivalent of $f(\G)$ gravity, holds for any $ \displaystyle n=\frac{d+1}{4}$\,, in agreement with the trace Eq. \eqref{treq}. For this reason, in four dimensions, it trivially provides a constant line element. In any case, the asymptotic flatness is always recovered in more than five dimensions; furthermore, solutions \eqref{secondND} and \eqref{firstND} admits as horizon $r_S \sim (GM)^{\frac{2}{d-4}}$.

Let us now see some more specific solutions of the system \eqref{ELP}-\eqref{ELG}, analyzing the boundary cases $d=3$ and $d=4$. In $d = 3$ we have the following solution for any $P(r)$
\begin{equation}
Q(r) = \frac{1}{3}\left(A(r) - e^{q_0}P'(r)+\frac{e^{2q_0}}{A(r)}P'(r)^2\right)\,,
\end{equation} 
 with 
$$A(r) = \left(\frac{27e^{q_0}}{2}P'(r)-e^{3q_0}P'(r)^3 +\frac{3e^{q_0}}{2}P'(r)\sqrt{81-12e^{2q_0}P'(r)^2}\right)^{1/3}\,.$$
$q_0$ is an integration constant and the Gauss-Bonnet term vanishes in this case. As an example, by introducing the relation $Q(r) = P(r)^k$, we find that the field equations are satisfied by any $P(r)$ solving the equation
\begin{equation}
k (P^{2 k}-3)P'^2 -  P(P^{2 k}-1) P''= 0.
\label{eqPQ}
\end{equation} 
The limit $k=-1$ provides back solution \eqref{firstND}; an interesting analytic solution of Eq. \eqref{eqPQ} occurs for $k = - 1/3$, where the components of the interval are:
\begin{eqnarray}
P(r)^2 = - 2 c_1 \left[(r+c_2) \left(\frac{6 r}{M(r)}+1 \right) \right]+\frac{3}{8} \left[\frac{M(r)^2 + 9}{M(r)} + 3 \right] \nonumber
\end{eqnarray}
with
\begin{equation}
M(r) = \sqrt[3]{128 c_1^2 r^2+16 \left(16 c_1 c_2-9\right) c_1 r+64
   \sqrt{c_1^3 \left(c_2+r\right){}^3 \left(4 c_1 r+4 c_2 c_1-1\right)}+128 c_2^2.
   c_1^2-144 c_2 c_1+27}.
\end{equation} 
Moreover, in $d=4$ we only get the following solutions for constant $\G$,
\begin{gather}\label{sol1}
P(r)^2 = -\frac{1}{2}\exp \left[tanh^{-1}\left(\sqrt{\frac{\G_0}{30}}\frac{r^2}{2}\right) \right]\sqrt{4-\frac{\G_0 r^4}{30}} \,\, \text{and}\,\, Q(r)^{-2} = 1 + \frac{\sqrt{\G_0}r^2}{2 \sqrt{30}}\,\, \text{for} \,\,n=5/4\,, \\  \label{sol3}
P(r)^2 = 1 = Q(r)^{2}\,\,, \text{for}\,\, \G _0 = 0 \,\, \text{and} \,\, \forall n\,.
\end{gather}
If we Taylor expand $P(r)^2$ in the first solution \eqref{sol1}, we find that $P(r)^2 = Q(r)^{-2}$, which is an AdS-like solution, where  $k^2 \equiv \frac{\sqrt{\G_0}}{2\sqrt{30}}$ can be considered as the bulk cosmological constant. The second one is Minkowski. In all of the above cases we set the integration constants so that we have the correct asymptotic behavior.

\section{Conclusions}
\label{secV}

Higher dimensional theories have been extensively studied to provide solutions to address some shortcomings of GR. Gauss-Bonnet gravity is one of them \cite{Charmousis:2002rc}. In this paper,  we studied  a generalized  Gauss-Bonnet gravity of the form of \eqref{f(G)}, in arbitrary $(d+1)$ dimensions. Specifically, using the Noether Symmetry Approach, we found forms of the function $f$, which are invariant under point transformations, in a spherically symmetric background. As it turns out, the only possible form is a power-law $f(\G) = f_0 \G^n$. In this perspective, the standard action of GR is recovered for $f(\G) = f_0 \G^{1/2}$ as soon as the degrees of freedom related to $R$ are dominant with respect to the others, like in the case of cosmology.
Furthermore, we considered the above power-law model in arbitrary $(d+1)$ dimensions, and found analytical static and spherically symmetric solutions of the form \eqref{sphsymmet}. The solutions we found are summarized in the following Table \ref{Table1}.

\begin{table}[H]
\centering
\resizebox{16cm}{!}{\begin{tabular}{ |c | c |c|c|} 
\hline
$\mathbf{P(r)^2}$ & $\mathbf{Q(r)^2}$ & $\mathbf{d}$ & $\mathbf{n}$ \\ \hline 
 $\displaystyle 1 + e^{-2 c_2} \sqrt{c_1-4 r} r^{\frac{3}{2}-\frac{d}{2}}$ & $1/P(r)^2$ & $d \ge 3$ & $n>0, \neq 1$ \\ \hline
 $\displaystyle P_0^2 \left(1- \frac{k_3}{r^{\frac{d}{2}-2}} \right)$ & $1/P(r)^2$ & $d>3$ & $n=1$ \\ \hline
 $\displaystyle 1 \pm r^{2-\frac{d}{2}}\sqrt{\frac{4 k_1 d}{120 \binom{d+1}{d-4} }} \pm r^2 \sqrt{ \frac{ \G_0 (d-3) }{120 \binom{d+1}{d-4} }}$ & $1/P(r)^2$ & $d>3$ & $n = \frac{d+1}{4}$ \\ \hline 
 $\forall P(r)$ & $\displaystyle \frac{1}{3}\left(A(r) - e^{q_0}P'(r)+\frac{e^{2q_0}}{A(r)}P'(r)^2\right)$ & $d=3$ & $n>0$ \\ \hline 
 $\displaystyle -\frac{1}{2}\exp \left[tanh^{-1}\left(\sqrt{\frac{\G_0}{30}}\frac{r^2}{2}\right) \right]\sqrt{4-\frac{\G_0 r^4}{30}}$ & $\displaystyle 1+ \frac{\sqrt{\G_0}r^2}{2\sqrt{30}}$ & $d=4$ & $n = 5/4$ \\ \hline
 $1$ & $1$ & $ d=4$ & $\forall n$ \\ \hline
  \end{tabular} }
\caption{Exact static and spherically symmetric solutions in $f(G)$ gravity, for $f(\G) = f_0 \G^n$ in arbitrary $d+1$ dimensions.}\label{Table1}
 \end{table}

In a future work, we will  study the stability of the above solutions, as well as the possibility to find some compact object solutions for models with specified $d$ and $n$. 

\section*{Acknowledgments}
F.~B and S.~C. acknowledge the support of  {\it Istituto Nazionale di Fisica Nucleare} (INFN) ({\it iniziative specifiche} MOONLIGHT2, GINGER and QGSKY). This paper is based upon work from COST action CA15117 (CANTATA), supported by COST (European Cooperation in Science and Technology).

\end{document}